\pdfoutput=1
\documentclass[aps,reprint,twocolumn,showpacs,superscriptaddress,nofootinbib]{revtex4-1}  
\usepackage{tabularx}
\makeatletter
\def\hlinewd#1{
\noalign{\ifnum0=`}\fi\hrule \@height #1 
\futurelet\reserved@a\@xhline}
\makeatother
\usepackage{scrextend}
\usepackage{amsmath}
\usepackage{amssymb}
\usepackage{epsfig}
\usepackage{graphicx}
\usepackage{hyperref}
\usepackage{dcolumn}  
\usepackage{slashed}
\usepackage{color}
\usepackage{rotating}
\usepackage[margin=0.9in,a4paper]{geometry}
\usepackage[table,xcdraw,dvipsnames]{xcolor}
\usepackage[utf8]{inputenc}
\usepackage{colortbl}
\usepackage[normalem]{ulem}
\usepackage{mathrsfs} 
\usepackage[version=4]{mhchem}
\usepackage{cancel}

\usepackage{tikz}
\usepackage{tikz-feynman}
\definecolor{darkgreen}{rgb}{0.0, 0.5, 0.0}

\usepackage{simpler-wick}

\newcommand{\Sec}[1]{ \medskip \noindent {\sl \bfseries #1}}

\usepackage{comment}

\definecolor{nicered}{rgb}{0.7,0.1,0.1}
\definecolor{nicegreen}{rgb}{0.1,0.5,0.1}
\definecolor{red}{rgb}{1.0, 0, 0}
\definecolor{niceblue}{rgb}{0,0,0.8}
\definecolor{red}{rgb}{1.0, 0, 0}
\hypersetup{colorlinks,citecolor= nicegreen,linkcolor= nicered,urlcolor=nicered}

\begin{document}

\title{Constraints on baryon-number-violating top-quark operators\\ in standard model effective field theory }

\author{Hector Gisbert}
\email[email: ]{hector.gisbertmullor@unipd.it}
\affiliation{Istituto Nazionale di Fisica Nucleare (INFN), Sezione di Padova, Via F. Marzolo 8, 35131 Padova, Italy}
\affiliation{Dipartimento di Fisica e Astronomia ‘G. Galilei’, Università di Padova, Via F. Marzolo 8, 35131 Padova, Italy}
\affiliation{Escuela de Ciencias, Ingenier\'ia y Dise\~{n}o, Universidad Europea de Valencia, \\
Passeig de la Petxina 2, 46008 Valencia, Spain}

\author{Antonio Rodríguez-Sánchez}
\email[email: ]{arodrigu@sissa.it}
\affiliation{SISSA, Via Bonomea 265, 34136 Trieste, Italy}
\affiliation{INFN, Sezione di Trieste, SISSA, Via Bonomea 265, 34136 Trieste, Italy}
\affiliation{Departament de F\'{i}sica Te\`{o}rica, Instituto de F\'{i}sica Corpuscular,\\
Universitat de Val\`encia -- Consejo Superior de Investigaciones Cient\'{i}ficas,\\
Parc Cient\'{i}fic, Catedr\'{a}tico Jos\'{e} Beltr\'{a}n 2, E-46980 Paterna, Valencia, Spain}
\author{Luiz Vale Silva}
\email[email: ]{luizva@ific.uv.es}
\affiliation{Departament de F\'{i}sica Te\`{o}rica, Instituto de F\'{i}sica Corpuscular,\\
Universitat de Val\`encia -- Consejo Superior de Investigaciones Cient\'{i}ficas,\\
Parc Cient\'{i}fic, Catedr\'{a}tico Jos\'{e} Beltr\'{a}n 2, E-46980 Paterna, Valencia, Spain}
\affiliation{Departamento de Matem\'{a}ticas, F\'{i}sica y Ciencias Tecnol\'{o}gicas,\\
Universidad Cardenal Herrera-CEU, CEU Universities,\\
46115 Alfara del Patriarca, Val\`{e}ncia, Spain}

\date{\today}

\begin{abstract}
Within the Standard Model Effective Field Theory framework, we set indirect constraints on top quark operators that violate baryon number by one unit above the TeV scale. We find that these constraints are typically many orders of magnitude more stringent than the recently 
derived direct bounds from collider experiments. Therefore, direct observation of baryon number violation in these top-quark observables at the TeV scale would imply a large fine-tuning among operators across different energy scales. This possibility is not protected under universal radiative corrections or any known symmetry principles.
\end{abstract}

\maketitle

\Sec{Introduction.}
The conservation of Baryon Number ($B$) is crucial for the stability of matter in the universe \cite{Weyl:1929fm}. However, Baryon Number Violation (BNV) is indispensable for explaining the observed baryon asymmetry, a necessary ingredient for successful baryogenesis \cite{Sakharov:1967dj}.
$B$ conservation, which appears as an accidental symmetry of the Standard Model (SM) at the perturbative level,
is  
broken by higher-dimensional operators built from the same dynamical fields and symmetry principles, suppressed by an unknown heavy scale 
\cite{Weinberg:1979sa}.

Grand Unified Theories
(GUTs)
suggest that BNV should manifest through processes such as proton decay, accommodating extraordinarily long proton lifetimes on the order of $10^{30}$ years or more. Experimental efforts, including those at underground detectors like Super-Kamiokande, actively search for proton decay, which would provide critical evidence in favor of these unified theories~\cite{Nath:2006ut}.

Expanding the scope of BNV studies beyond traditional searches, recent research has explored interactions involving heavier fermions. Ref.~\cite{Hou:2005iu} examines operators involving third-generation flavors. Tau-related BNV processes are discussed in Refs.~\cite{Hou:2005iu,Crivellin:2023ter,Heeck:2024jei}, while BNV effects in bottom physics are explored in Ref.~\cite{Beneke:2024hox,footnote}. Loop effects involving higher dimensional operators with different flavor structures are studied in Ref.~\cite{Gargalionis:2024nij}. Possible signatures in colliders of BNV processes are also discussed in Ref.~\cite{Morrissey:2005uza}.

In the top-quark sector, a study by the CMS collaboration using proton-proton collision data at $\sqrt{s} = 13$ TeV and an integrated luminosity of 138 fb$^{-1}$ investigated BNV in top quark production and decay processes~\cite{PhysRevLett.132.241802}. This study provides the most stringent direct constraints to date on BNV top quark operators, probing Wilson coefficients as small as $ 0.7 - 0.02$~TeV$^{-2}$ depending on the fermion flavor combination, thus  improving previous direct experimental bounds on BNV branching fractions \cite{CMS:2013zol} by three to six orders of magnitude.

The top quark stands out among SM fermions due to its heavy mass. Beyond tree level, the top-quark Yukawa coupling $y_t$ has motivated numerous Standard Model Effective Field Theory (SMEFT) studies that highlight top-quark contributions, especially involving interactions conserving baryon number \cite{Fox:2007in,Grzadkowski:2008mf,Drobnak:2011aa,Brod:2014hsa,Altmannshofer:2023bfk,Cirigliano:2016nyn,Bissmann:2019gfc,Aguilar-Saavedra:2018ksv,Maltoni:2019aot,Brivio:2019ius,Durieux:2019rbz,Hartland:2019bjb,Bissmann:2020mfi,Bruggisser:2021duo,Ethier:2021bye,Miralles:2021dyw,Durieux:2022cvf,deBlas:2022ofj,Bruggisser:2022rhb,Giani:2023gfq,Kassabov:2023hbm,Grunwald:2023nli,Garosi:2023yxg,Endo:2018gdn,Aebischer:2020lsx,ValeSilva:2022tph}. Regarding BNV, Ref.~\cite{Dong:2011rh} not only focuses on direct LHC bounds, but also highlights the tree-level contributions from $udt\ell$ operators, with $\ell = e, \mu$, where $W$ exchange drives BNV nucleon decay.
Ref.~\cite{Beneito:2023xbk} discusses BNV mediated by dimension-7 operators involving light flavors, which receive a $y_t^2$ enhancement through top-loop contributions to the wave function renormalization of the scalar leg~\cite{Liao:2016hru}. Alternative scenarios involving higher dimensional operators beyond $\Delta B=\Delta L=1$ have been explored in Refs.~\cite{Durieux:2012gj,Durieux:2013uqa}.

In this work, we use the SMEFT framework to derive model-independent
limits on dimension-6 top quark BNV operators from nucleon decay lifetimes. We provide a comprehensive analysis, identifying the internal mechanisms that induce low-energy BNV processes from top quark operators and find that constraints from nucleon lifetimes are typically 20 orders of magnitude more stringent than direct collider limits~\cite{PhysRevLett.132.241802}. Thus, our results suggest that directly observing BNV in these top-quark processes at the TeV scale through collider experiments would generally require extreme fine-tuning among operators across different low-energy processes and energy scales.

\Sec{Low-energy observables.} The observables considered in our analysis are the nucleon lifetimes for decays where a nucleon ($N = p, n$) decays into a pseudoscalar ($P = \pi,\, K,\, \dots$) and a lepton ($\mathcal{L} = \ell,\,\nu_\ell$ with $\ell = e, \mu$), indicating $\Delta B = 1$ BNV. These decays provide the most stringent and theoretically clean constraints. We summarize them in Table~\ref{tab:nucleon_lifetimes}.

\begin{table}[h!]
    \centering
     \centering
 \resizebox{0.25\textwidth}{!}{
    \begin{tabular}{c|c}
        \hline
  \hline
  Channel & Limit [$10^{30}$ years] \\
        \hline
  \hline
        $p \to \pi^0 e^+$ & $2.4 \times 10^4$~\cite{Super-Kamiokande:2020wjk}  \\
        $p \to \pi^0 \mu^+$ & $1.6 \times 10^4$~\cite{Super-Kamiokande:2020wjk}  \\
        $p \to \pi^+ \bar{\nu}$ & $3.9 \times 10^2$~\cite{Super-Kamiokande:2013rwg}   \\
        $p \to K^0 e^+$ & $1.0 \times 10^3$~\cite{Super-Kamiokande:2005lev}   \\
        $p \to K^0 \mu^+$ & $4.5 \times 10^3$~\cite{Super-Kamiokande:2022egr}   \\
        $p \to K^+ \bar{\nu}$ & $5.9 \times 10^3$~\cite{Super-Kamiokande:2014otb}   \\
        \hline
        $n \to \pi^- e^+$ & $5.3 \times 10^3$~\cite{Super-Kamiokande:2017gev}   \\
        $n \to \pi^- \mu^+$ & $3.5 \times 10^3$~\cite{Super-Kamiokande:2017gev}   \\
        $n \to \pi^0 \bar{\nu}$ & $1.1 \times 10^3$~\cite{Super-Kamiokande:2013rwg}   \\
        $n \to K^0 \bar{\nu}$ & $1.3 \times 10^2$~\cite{Super-Kamiokande:2005lev}  \\
        \hline
    \end{tabular}}
    \caption{Limits on nucleon lifetimes for various decay channels (all at 90\% C.L.).    }
    \label{tab:nucleon_lifetimes}
\end{table} 

The low-energy description of these decays is given by the Low-energy Effective Field Theory (LEFT), where the heavy SM degrees of freedom have been integrated out. In the current state-of-the-art, the running and matching of dimension-6 LEFT operators are known up to one-loop order~\cite{Jenkins:2017dyc,Jenkins:2017jig,Dekens:2019ept}, and the branching ratios in this basis can be readily obtained using the results from Ref.~\cite{Nath:2006ut} (see also Ref.~\cite{Beneito:2023xbk} for $p\to K^0 \ell^+$).

\Sec{From top operators to light-quark transitions.} Given the energy scale of nucleon decay, the LEFT provides a complete description of the corresponding BNV processes, assuming they are induced by heavy particles, such as leptoquarks~\cite{Dorsner:2012nq}. Top quarks are not active degrees of freedom at low energies and, consequently, there are no top-quark operators in the LEFT framework. However, the presence of top quarks, like any other high-energy particle, leaves a measurable imprint on low-energy dynamics, and operators involving top quarks are no exception.

Assuming BNV is induced by particles above the electroweak scale, these effects can be studied in a model-independent manner using the SMEFT framework, including all dimension $D > 4$ operators with SM fields that preserve the SM gauge symmetries, $SU(3)_C \times SU(2)_L \times U(1)_Y$. In this context, one might consider top-quark operators such as $\varepsilon_{\alpha\beta\gamma}\, d_{R}^\alpha C u_{R}^{\beta}\; t_{L}^{\gamma} C e_L$, where $C$ denotes the Dirac matrix for charge conjugation and the Greek letters indicate $SU(3)_C$ color indices. However, this operator is not invariant under $SU(2)_L$ and is therefore not a valid SMEFT operator. To achieve gauge invariance, it must be extended to $\varepsilon_{\alpha\beta\gamma}\, d_{R}^\alpha C u_{R}^{\beta} \; ( t_{L}^{\gamma} C e_L - b_{L}^{\gamma} C \nu_L)$, naturally connecting processes with and without tops in a model-independent way. A complete basis of $\Delta B = 1$ BNV dimension-6 SMEFT operators is given in Refs.~\cite{Abbott:1980zj,Alonso:2014zka},
\begin{align}
\mathcal{Q}_{prst}^{duq\ell} &= \varepsilon_{\alpha\beta\gamma}\,\varepsilon_{ij}\,(d_p^\alpha\,C\,u_r^\beta)\,(q_s^{i\gamma}\,C\,\ell_t^j)~,\label{eq:BNVSMEFTop1}\\
\mathcal{Q}_{prst}^{qque} &= \varepsilon_{\alpha\beta\gamma}\,\varepsilon_{ij}\,(q^{i\alpha}_p\,C\,q_r^{j\beta})\,(u^{\gamma}_s\,C\,e_t)~,\label{eq:BNVSMEFTop2}\\
\mathcal{Q}_{prst}^{qqq\ell} &= \varepsilon_{\alpha\beta\gamma}\,\varepsilon_{il}\,\varepsilon_{jk}\,(q_p^{i\alpha}\,C\,q_r^{j\beta})\,(q_s^{k\gamma}\,C\,\ell_t^l)~,\label{eq:BNVSMEFTop3}\\
\mathcal{Q}_{prst}^{duue} &= \varepsilon_{\alpha\beta\gamma}\,(d^\alpha_p\,C\,u_r^\beta)\,(u^\gamma_s\,C\,e_t)~.\label{eq:BNVSMEFTop4}
\end{align}
Here, $q$ and $\ell$ represent the left-handed doublets of quarks and leptons, respectively, with $q = (u_L, d_L)^T$ and $\ell = (\nu_L, e_L)^T$. The symbols $u$, $d$, and $e$ are used for the right-handed fermions corresponding to up-type quarks, down-type quarks, and charged leptons, respectively. Roman letters $i$ to $l$ refer to $SU(2)_L$ indices, while $p$ to $t$ denote flavor (generation) indices ranging from 1 to 3. These operators~\eqref{eq:BNVSMEFTop1}-\eqref{eq:BNVSMEFTop4} form a closed set under the renormalization group equations (RGEs)~\cite{Alonso:2014zka}.

The nucleon decays introduced previously are mediated by light-quark LEFT operators in the mass basis, $u'{}_{L,R}^{1}$ and $d'{}_{L,R}^{1,2}$. However, the quark fields in SMEFT are defined in the flavor basis, where the Yukawa matrices are not diagonal. In a general setup, these bases are related by a unitary transformation in flavor (generation) space,
\begin{equation}
u_{L}= U_{L}^u u'_{L} \, ,\quad d_{L}= U_{L}^d d'_{L} \, ,  
\end{equation}
and analogous ones for the right-handed fields. In our previous example,
\begin{align}\nonumber
\mathcal{Q}_{113\ell}^{duq\ell}&=\varepsilon_{\alpha\beta\gamma}\, (d_{R}^{1 \alpha}\,C\,u_{R}^{1 \beta}) ( u_{L}^{3 \gamma} C e_{L,\ell} -  d_{L}^{3 \gamma} C \nu_{L,\ell})
\\&
\nonumber
\supset \varepsilon_{\alpha\beta\gamma}\;  U_{R,11}^u U_{R,11}^d \; (d'{}_{R}^{1 \alpha}\,C\,u'{}_{R}^{1 \beta}) \\
& \cdot ( U_{L,31}^u \, u'{}_{L}^{1 \gamma} C e_{L,\ell} - U_{L,31}^d \, d'{}_{L}^{1 \gamma} C \nu_{L,\ell}) \, .
\end{align}
Due to the left-handed fields being part of $SU(2)_L$ doublets, it is not possible to simultaneously diagonalize both $Y_u$ and $Y_d$. Consequently, one cannot choose a flavor basis where both $u_L^p=u'{}_L^p$ and $d_L^p=d'{}_L^p$ hold simultaneously. Therefore, as illustrated in the example above, top-quark operators with $t_L \subset q^3$ in any flavor basis will generally induce light-quark operators in the mass basis at tree level, leading to nucleon decay.

There is, however, the freedom to select a flavor basis in which $Y_u$ is diagonal. In such a basis, one might focus on operators involving only $t_R=t'_R$, to avoid the induction of light quarks at tree level. The first important caveat is the instability of the diagonality of $Y_u$ under SM running, as emphasized in Ref.~\cite{Aebischer:2020lsx}. If nature is selective enough to couple only to $t_R$ in a basis where $Y_u$ is diagonal, it appears more natural for this to occur at a high UV scale $\Lambda_\text{UV}$. However, even if that were the case, the up-type Yukawa matrix will no longer be diagonal at lower energy scales, such as the electroweak scale $\Lambda_\text{EW}$, due to contributions from the non-diagonal $Y_d$ matrix. The back rotation required to re-diagonalize $Y_u$ will then inevitably induce light-quark operators. Analytically, one finds for $t_R$ (analogously for $c_R$), in the leading-log approximation,
\begin{equation}\label{eq:backrotation}
t_R  \to  -\frac{3}{2}\frac{\epsilon_\pi\,y_t \, y_u}{y_t^2-y_u^2}\ln\frac{\Lambda_\text{UV}^2}{\Lambda_\text{EW}^2} \sum_{k=d,s,b}V_{tk} \, y_k^2 \, V_{uk}^*  \,  u'_R \, , 
\end{equation}
where $V$ is the Cabibbo-Kobayashi-Maskawa (CKM) matrix and $\epsilon_\pi \equiv 1/(4\pi)^2$; see the supplementary material for its derivation. This effect can still be evaded if the UV model is defined in a basis where the Yukawa matrix is diagonal at $\Lambda_\text{EW}$. From this point onward, we will work under this assumption.

\begin{figure}[t]
    \centering
\begin{tikzpicture}
        \begin{feynman}
            \vertex (a) at (0, 0);
            \vertex [above left=1.5cm and 1.5cm of a] (b);
            \vertex [above right=1.5cm and 1.5cm of a] (c);
            \vertex [below left=0.75cm and 0.75cm of a] (d);
            \vertex [below right=0.75cm and 0.75cm of a] (e);
            \vertex [below left=0.75cm and 0.75cm of d] (f2);
            \vertex [below right=0.75cm and 0.75cm of e] (f1);

            \diagram* {
                (a) -- [fermion] (b),
                (a) -- [fermion] (c),
                (a) -- [fermion,blue] (d),
                (a) -- [fermion] (e),
                (d) -- [scalar]  (e),
                (d) -- [fermion] (f2),
                (e) -- [fermion] (f1),
            };

            \vertex at (a) {\(\blacksquare\)};
            \fill[red] (-0.75,-0.75) circle (3pt);
            \fill[red] (0.75,-0.75) circle (3pt);
            \node at (2,0) ;
        \end{feynman}
    \end{tikzpicture}
    \caption{\label{fig:higgsexchange}An example of Feynman diagram illustrating the SMEFT running of a BNV operator (black square vertex). The red vertices denote the Yukawa couplings, connected by the dashed line representing the Higgs field. The black lines correspond to light quarks, while the blue line represents a top quark, which is converted into a light quark via Yukawa insertions.}
\end{figure}
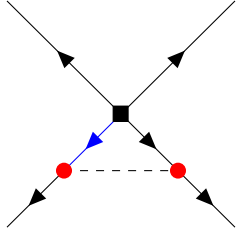

We now turn to the second important caveat. From the perspective of the flavor-to-mass basis rotation, the setup described above avoids the introduction of light quarks in the first approximation. However, pure SM interactions contain mechanisms that convert $t_R$ into light quarks, regardless of the choice of basis. As a result, interactions involving only light-quark generations inevitably emerge if an operator with $t_R$ is introduced at a high energy scale. Indeed, using the SMEFT running computed in Ref.~\cite{Alonso:2014zka}, the corresponding top-philic setup is not stable under the universal
running effects of BNV operators. The primary
mechanism driving operator mixing involves Higgs exchanges, as illustrated in Fig.~\ref{fig:higgsexchange}. 

As an example, $C^{duq\ell}_{131\ell}$ appears in the $\beta$ function of $C^{qqq\ell}_{131\ell}$. Thus, even if $C^{qqq\ell}_{131\ell}(\Lambda_\text{UV}) = 0$,
\begin{equation}\label{eq:example_single_Log}
C^{qqq\ell}_{131\ell}(\Lambda_{\mathrm{EW}})\propto C^{duq\ell}_{131\ell}(\Lambda_{\mathrm{UV}})\,\epsilon_\pi\ln\frac{\Lambda_{\mathrm{UV}}^2}{\Lambda_{\mathrm{EW}}^2} + \cdots \, ,
\end{equation}
BNV effects are consequently induced in tree-level nucleon decays. In this particular example, a third-generation $SU(2)_L$ doublet is generated at $\Lambda_\text{EW}$, which necessarily includes light flavors. We have verified that most operators not directly probed at tree level are tested through this mechanism. The few remaining operators produce effects via one-loop $\beta$ functions, arising from terms of the form $\left(\epsilon_{\pi} \ln\Lambda_{\mathrm{UV}}^2/\Lambda_{\mathrm{EW}}^2\right)^2$ rather than the single logarithm in Eq.~\eqref{eq:example_single_Log}, i.e., through a chain of operator mixing, each link being generated at one loop.

The model-independent mechanisms described here inducing nucleon decays carry some suppression, either due to small CKM angles in the flavor-to-mass basis rotation, loop factors, or small Yukawa couplings. However, 
this suppression is largely overcompensated by the superior sensitivity of nucleon decay experiments compared to direct BNV searches at the LHC. As a result, indirect bounds on top-quark operators are going to be typically many orders of magnitude stronger than the direct searches.

\Sec{Bounds on top BNV SMEFT operators}. In this section, we present an analysis of indirect bounds on top-quark operators. As motivated above, we work in the basis in which the up quark Yukawa matrix is diagonal at $\Lambda_\text{EW}$, so that $q = (u_L, V d_L)^T$. In this basis, we generate the top operators at the UV scale, $\Lambda_\text{UV} = 1\, \mathrm{TeV}$. Our analysis includes not only the key effects discussed above but also the full dimension-6 SMEFT one-loop running from the TeV scale down to the electroweak  scale~\cite{Alonso:2014zka}, the one-loop matching to the dimension-6 LEFT~\cite{Dekens:2019ept}, and the LEFT running~\cite{Jenkins:2017dyc} down to the low-energy scale. In practice, we use the \texttt{Mathematica} package \texttt{DsixTools}~\cite{Celis:2017hod,Fuentes-Martin:2020zaz}. Additional technical details on our specific setup are provided in the supplementary material.

Once the branching ratios are obtained as functions of the SMEFT Wilson coefficients at the TeV scale, we fit them to the current experimental limits in Table~\ref{tab:nucleon_lifetimes}, including one Wilson coefficient at a time. In the few cases where redundancies arise, such as $\mathcal{Q}^{qque}_{1311} = \mathcal{Q}^{qque}_{3111}$, we adopt the convention of inducing the corresponding Wilson coefficients with the same strength. For instance, in this example, we display only the upper experimental bound on $C^{qque}_{1311}$, assuming that in the original $1\, \mathrm{TeV}$ Lagrangian, all Wilson coefficients are zero except for $C^{qque}_{3111} = C^{qque}_{1311}$. Once again we relegate further details to the supplementary material.

\begin{figure*}[tbh]
    \centering
    \includegraphics[width=\textwidth]{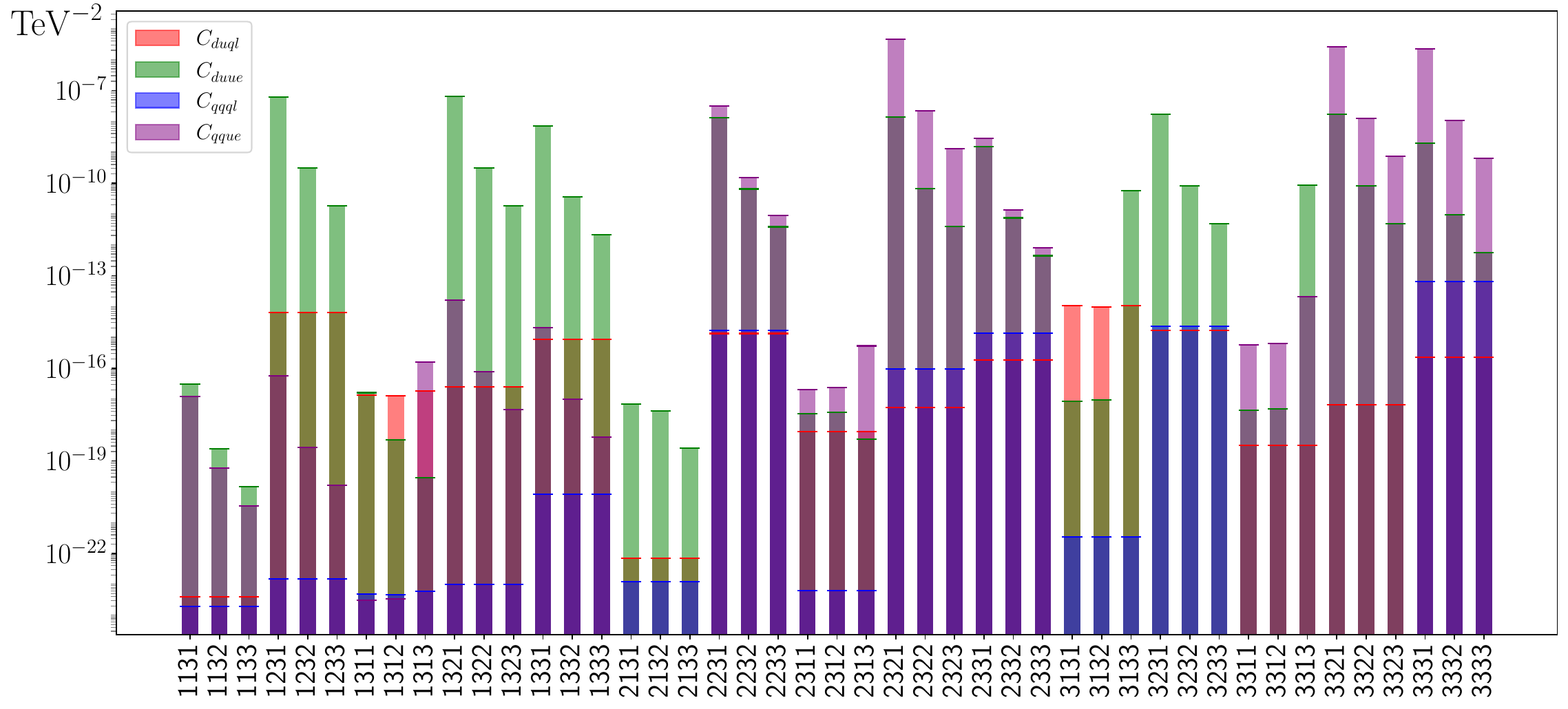}
    \caption{Constraints on $\Delta B=1$ SMEFT Wilson coefficients~\eqref{eq:BNVSMEFTop1}-\eqref{eq:BNVSMEFTop4} (red, purple, blue, and green, respectively) at 68\% C.L., in TeV$^{-2}$ units, derived from limits on $N\to P\,\bar{\mathcal{L}}$ nucleon lifetimes, see Table~\ref{tab:nucleon_lifetimes}.
    }
    \label{fig:topBNV_bounds}
\end{figure*}

The experimental limits on all Wilson coefficients associated with top operators are displayed in Fig.~\ref{fig:topBNV_bounds}. In all cases, these bounds are several orders of magnitude more stringent than the corresponding direct limits from Ref.~\cite{PhysRevLett.132.241802}.

\Sec{Hierarchy of constraints on top-quark operators.}
To better understand the hierarchy of the bounds, we have repeated the analysis, but retaining only the tree-level SMEFT-LEFT matching factors and the leading-log effects. At this stage, a comment is in order. In a new physics scenario where UV couplings are of order one, with $\Lambda_{\mathrm{BNV}} \sim 10^{13}\, \mathrm{TeV}$, truncating the leading logarithmic series (scaling as powers of $\epsilon_{\pi} \ln\Lambda_{\mathrm{UV}}^2/\Lambda_{\mathrm{EW}}^2$ multiplied by order one couplings such as the gauge and top Yukawas) may be unjustified. In such cases, one would need either to perform the full one-loop RGE running, as we have implemented numerically above, or to solve it analytically, resumming at least the problematic logarithmic series with order one couplings. However, within our setup, where top operators are generated at the TeV scale, the logarithms are large but truncating the series remains a safe approximation.

\begin{table*}[tbh]
    \centering
    \renewcommand{\arraystretch}{1.4} 
    \begin{tabular}{|c|c|c||c|c|c|}
        \hline
        \multicolumn{3}{|c||}{ $\boldsymbol{C^{duq\ell}}$} & \multicolumn{3}{c|}{$\boldsymbol{C^{qque}}$} \\
        \cline{1-3} \cline{4-6}
        $1 \, 1 \, 3 \, \ell$ & $V_{32}$ & $p\to K^+ \bar{\nu}$ & $1 \, 3 \, 1 \, 1$ & $V_{31}$ & $p\to \pi^0 e^+$ \\
        \cline{1-3} \cline{4-6}
        $2 \, 1 \, 3 \, \ell$ & $V_{31}$ & $p\to K^+ \bar{\nu}$ & $1 \, 3 \, 1 \, 2$ & $V_{31}$ & $p\to \pi^0 \mu^+$ \\
        \cline{1-3} \cline{4-6}
        $i \, a \, 3 \, \ell$ & $(Y_d)_{1i} \, (Y_u)_{aa} \, V_{a1} \, V_{32} \, L$ & $p\to K^+ \bar{\nu}$ & $i \, 3 \, 1 \, 3$ & $(Y_e)_{33} \, (Y_d)_{32}  \, V_{i1} \, L$ & $p\to K^+ \bar{\nu}$ \\
        \hline
        $i \, 3 \, 2 \, \ell$ & $(Y_d)_{1i} \, (Y_u)_{33} \, V_{31} \, V_{22} \, L$ & $p\to K^+ \bar{\nu}$ & $a \, 3 \, 1 \, 1$ & $(Y_d)_{13} \, (Y_d)_{33} \, V_{a1} \, L$ & $p\to \pi^0 e^+$ \\
        \hline
        $1 \, 3 \, 1 \, \ell$ & $(Y_d)_{11} \, (Y_u)_{33} \, V_{11} \, V_{32} \, L$ & $p\to K^+ \bar{\nu}$ & $a \, 3 \, 1 \, 2$ & $(Y_d)_{13} \, (Y_d)_{33} \, V_{a1} \, L$ & $p\to \pi^0 \mu^+$  \\
        \hline
        $a \, 3 \, 1 \, \ell$ & $(Y_d)_{2a} \, (Y_u)_{33} \, V_{22} \, V_{31} \, L$ & $p\to K^+ \bar{\nu}$ & $1 \, 3 \, 2 \, \ell$ & $(Y_e)_{\ell\ell} \, (Y_u)_{22} \, V_{22} \, V_{31} \, L$ & $p\to K^+ \bar{\nu}$ \\
        \hline
        $3 \, 1 \, 3 \, \ell$~$^*$ & $(Y_d)_{23} \, (Y_d)_{32} \, V_{21} \, L$ & $p\to K^+ \bar{\nu}$ & $1 \, i \, 3 \, \ell$ & $(Y_e)_{\ell\ell} \, (Y_u)_{33} \, V_{i1} \, V_{32} \, L$ & $p\to K^+ \bar{\nu}$  \\
        \hline
        \hline
        \multicolumn{3}{|c||}{$\boldsymbol{C^{qqq\ell}}$} & \multicolumn{3}{c|}{$\boldsymbol{C^{duue}}$} \\
        \hline
        $i \, 1 \, 3 \, \ell$ & $V_{i1} \, V_{32}$ & $p\to K^+ \bar{\nu}$ & $1 \, 1 \, 3 \, 1$ & $(Y_d)_{11} \, (Y_u)_{33} \, V_{31} \, L$ & $p\to \pi^0 e^+$ \\
        \hline
        $1 \, 3 \, 1 \, \ell$ & $V_{11} \, V_{32}$ & $p\to K^+ \bar{\nu}$ & $1 \, 1 \, 3 \, a$ & $(Y_e)_{aa} \, (Y_u)_{33} \, V_{32} \, L$ & $p\to K^+ \bar{\nu}$ \\
        \hline
        $2 \, 3 \, 1 \, \ell$ & $V_{22} \, V_{31}$ & $p\to K^+ \bar{\nu}$ & $1 \, 3 \, 1 \, 1$ & $(Y_d)_{11} \, (Y_u)_{33} \, V_{31} \, L$ & $p\to \pi^0 e^+$ \\
        \hline
        $1 \, 2 \, 3 \, \ell$ & $V_{21} \, V_{32}$ & $p\to K^+ \bar{\nu}$ & $1 \, 3 \, 1 \, a$ & $(Y_e)_{aa} \, (Y_u)_{33} \, V_{32} \, L$ & $p\to K^+ \bar{\nu}$ \\
        \hline
        $1 \, 3 \, 2 \, \ell$ & $V_{31} \, V_{22}$ & $p\to K^+ \bar{\nu}$ & $2 \, 1 \, 3 \, 1$ & $(Y_d)_{12} \, (Y_u)_{33} \, V_{31} \, L$ & $p\to \pi^0 e^+$ \\
        \hline
        $1 \, 3 \, 3 \, \ell$ & $g^2 \, V_{31} \, V_{32} \, L$ & $p\to K^+ \bar{\nu}$ & $2 \, 1 \, 3 \, a$ & $(Y_e)_{aa} \, (Y_u)_{33} \, V_{31} \, L$ & $p\to K^+ \bar{\nu}$ \\
        \hline
        $a \, 3 \, 3 \, \ell$ & $(Y_d)_{13} \, (Y_d)_{33} \, V_{a2} \, V_{31} \, L$ & $p\to K^+ \bar{\nu}$ & $2 \, 3 \, 1 \, c$ & $(Y_d)_{12} \, (Y_u)_{33} \, V_{31} \, L$ & $p\to \pi^0 \ell^+$ \\
        \hline
        $2 \, 2 \, 3 \, \ell$ & $(Y_d)_{23} \, (Y_d)_{13} \, V_{21} \, V_{32} \, L$ & $p\to K^+ \bar{\nu}$ & $2 \, 3 \, 1 \, 3$ & $(Y_e)_{33} \, (Y_u)_{33} \, V_{31} \, L$ & $p\to K^+ \bar{\nu}$ \\
        \hline
        $3 \, 2 \, 3 \, \ell$ & $(Y_d)_{33} \, (Y_d)_{13} \, V_{21} \, V_{32} \, L$ & $p\to K^+ \bar{\nu}$ & $3 \, 1 \, 3 \, c$ & $(Y_d)_{13} \, (Y_u)_{33} \, V_{31} \, L$ & $p\to \pi^0 \ell^+$ \\
        \hline
        $2 \, 3 \, 2 \, \ell$ & $(Y_d)_{33} \, (Y_d)_{13} \, V_{21} \, V_{22} \, L$ & $p\to K^+ \bar{\nu}$ & $3 \, 3 \, 1 \, c$ & $(Y_d)_{13} \, (Y_u)_{33} \, V_{31}  \, L$ & $p\to \pi^0 \ell^+$ \\
        \hline
    \end{tabular}
        \caption{Parametric suppression of leading effects constraining top-quark BNV operators for different Wilson coefficients at $\Lambda_{\mathrm{UV}}$. The table is divided into four blocks, each corresponding to a specific Wilson coefficient: $C^{duq\ell}$, $C^{qqq\ell}$, $C^{qque}$, and $C^{duue}$, respectively. For each set, the relevant indices ($c=1,2$~; $a=2,3$~; $i,\ell=1,2,3$) and parametric suppression ($L\equiv\epsilon_\pi\ln\Lambda_{\mathrm{UV}}^2/\Lambda_{\mathrm{EW}}^2$) are displayed in the first and second columns; $g$ corresponds to the electroweak coupling of the SM. The most constraining nucleon decay channels are indicated in the last column. $^*\,$Note that $C^{duq\ell}_{3132}$ is dominated by the $(Y_{d})_{13}\, (Y_{e})_{22}\, V_{31}\, L$ prefactor entering the experimental $p\to \pi^0 \mu^+$ channel.
        }
    \label{tab:parametric_suppression}
\end{table*}

Within the setup described above, we determine the parametric suppression of the prefactor that numerically dominates the experimental bound for each Wilson coefficient, which for simplicity we assume to be real at $\Lambda_{\mathrm{UV}}=1\, \mathrm{TeV}$. The corresponding results are presented in Table~\ref{tab:parametric_suppression}. Notice that $V_{31} (Y_{d,31})$ and $V_{13} (Y_{d,31})$ include a non-negligible CP-violating phase, which we take into account. Using this table, the overall hierarchy in Fig.~\ref{fig:topBNV_bounds} becomes clearer. Bounds on operators appearing at the leading-log level and involving light Yukawa insertions are relatively less stringent. When lepton Yukawa insertions are involved in the effect dominating the bound, operators with heavy leptons are more strongly constrained. Finally, the few Wilson coefficients that do not appear at order $\epsilon_\pi \ln\Lambda_{\mathrm{UV}}^2/\Lambda_{\mathrm{EW}}^2$ are the least constrained within the mechanisms studied here, since they only emerge at order $\left(\epsilon_\pi \ln\Lambda_{\mathrm{UV}}^2/\Lambda_{\mathrm{EW}}^2\right)^2$. Notice also how the few Wilson coefficients that do not appear at order $\epsilon_\pi \ln\Lambda_{\mathrm{UV}}^2/\Lambda_{\mathrm{EW}}^2$ are the least constrained within the mechanisms studied here, since they only emerge at order $\left(\epsilon_\pi \ln\Lambda_{\mathrm{UV}}^2/\Lambda_{\mathrm{EW}}^2\right)^2$. This also helps to identify cases where generally subleading effects, not considered here, could become relevant or even dominant. For instance, this applies to unknown loop contributions in the matching involving higher-dimensional effects in the LEFT framework (sometimes only suppressed by additional light Yukawa insertions) or unaccounted higher-order threshold effects in the decoupling of different quark flavors. 
These contributions are not enhanced by large $\ln\Lambda_{\mathrm{UV}}^2/\Lambda_{\mathrm{EW}}^2$ factors and therefore exhibit a different functional dependence compared to the effects considered here and thus no strong cancellations can be expected.
Given the relevance of the decay modes $ p \to K^+ \bar{\nu} $ and $ p \to \pi^0 \ell^+ $
in our study, a comment regarding future projections is in order. The bounds on their branching ratios are expected to improve by a factor $ 2 $ to $ 5 $
depending on the experiment~\cite{Hyper-Kamiokande:2018ofw,Theia:2019non,DUNE:2020ypp,JUNO:2022qgr} (see also Ref.~\cite{Domingo:2024qoj}). This improvement will enhance the sensitivity
shown in the Fig.~\ref{fig:topBNV_bounds} 
by a factor $ \sim 2 $.

\Sec{Discussion and conclusions.} Violation of Baryon Number has not yet been observed. Direct searches for BNV involving top operators at colliders, such as those studied in Ref.~\cite{PhysRevLett.132.241802}, have achieved remarkable precision and can be regarded as rigorous tests of SM symmetries. However, a different question is whether a nonzero result is expected in a general BSM scenario, considering the stringent low-energy bounds from nucleon decay lifetimes. The purpose of this work has been to explore this question in a model-independent manner.

Lagrangians with operators of dimension greater than four naturally arise in generic SM extensions once the heavy degrees of freedom are integrated out. These extensions are generally described by the SMEFT framework, which includes operators involving top quarks. On the other hand, nucleon decays are low-energy processes governed by the LEFT, which does not include top-quark operators. However, the presence of top-quark operators in the SMEFT Lagrangian induces LEFT operators with some degree of suppression.
We take the flavor basis in which
the up Yukawa matrix is diagonal at the electroweak
scale, thus avoiding tree-level conversion of $t_R$ into light quarks.
Taking the same up-type Yukawa matrix diagonal at 1~TeV, or taking the new physics scale as large as 100~TeV, would only change our estimated bounds by $\mathcal{O}(1)$ factors. On the other hand, choosing a much higher scale would lead to even stronger indirect bounds.
Within this setup, we conducted a comprehensive analysis using the current state-of-the-art methods to evaluate the low-energy impact of these operators, 
originated from new-physics dynamics
at the TeV scale. Our findings show that the flavor-to-mass basis rotation, combined with one-loop SMEFT running effects (which are universal and thus independent of the UV dynamics), is sufficient to set experimental bounds on all top-quark operators. These bounds are typically 20 orders of magnitude more constraining than those obtained from direct searches.

Our results should not be interpreted as a definitive no-go theorem, as we have not conducted (and cannot conduct, given current experimental and especially theoretical limitations) a global fit that accounts for all possible operators and effects. In this context, redundancy plays an important role, and precision tests of fundamental SM symmetries, such as those in Ref.~\cite{PhysRevLett.132.241802}, are always valuable.

However, the level of fine-tuning required to avoid the nucleon decay bounds studied here is extremely difficult to conceive, particularly given the absence of stringent symmetries preventing heavy-to-light quark conversion. For instance, one might attempt to induce a light-quark operator at the TeV scale that precisely cancels the effect of the top-quark operator in the low-energy observable. Achieving this would typically require a cancellation at the level of $10-20$ decimal points among the separate contributions of different Wilson coefficients, involving different infrared logarithms between the arbitrary TeV scale and the effective one corresponding to the low-energy process. Even if such a cancellation could be arranged at tree or one-loop level, it would be completely disrupted when incorporating the next order of corrections. In this context, the fine-tuning required for such a cancellation is, in practice, impossible to calculate and, even if it were to occur for a specific low-energy BNV observable, it would generally not hold across different processes.

Thus, our work has outlined the challenges in conceiving any BSM scenario with nonzero top-quark operators at the TeV scale that could be observed in collider searches. We have done so in a transparent manner, aiming to inspire potential alternative approaches to overcome these challenges.
For instance, promising avenues are exotic higher-order operators violating baryon and/or lepton number by more than one unit \cite{Durieux:2012gj} or the exploration of BNV operators in SM extensions involving long-lived particles
which cannot be produced in nucleon decays due to kinematic constraints \cite{Fornal:2018eol,Fox:2024kda,Li:2024liy}. However, such exploration lies beyond the scope of this work.

\Sec{Acknowledgments.}
We thank Arnau Bas i Beneito, Svjetlana Fajfer and Antonio Pich for valuable comments on the manuscript.
This work is supported by the European Union – Next Generation EU, by the University of Padua under the 2021 STARS Grants@Unipd programme (Acronym and title of the project: CPV-Axion – Discovering the CP-violating axion), by MIUR grant PRIN 2017L5W2PT, by MCIN/AEI/10.13039/501100011033 Grant No. PID2020-114473GB-I00, by the plan GenT program (CIDEIG/2023/12), by CSIC through IF/PF-ERC IFERC22013 and by PROMETEO/2021/071 (GV).

\appendix

\section{Analytic back rotation in the leading-log approximation}

At a particular renormalization scale $\Lambda$, one can work in a flavor basis in which $Y_u=D_u^0$ is diagonal, and $Y_d$ can be written as the product of a unitary matrix (the CKM matrix) and a diagonal one, i.e., $Y_d=V D_d^0$. However, this form is unstable under RGEs even when only considering dimension-4 interactions. As discussed in Ref.~\cite{Aebischer:2020lsx} (see also Ref.~\cite{Coy:2019rfr} for the analogous discussion when the Yukawa matrix $Y_d$ is diagonal at the scale $\Lambda$), one finds:
\begin{equation}
\mu^2\frac{dY_u}{d\mu^2}=\epsilon_\pi\, H_u Y_u \; ,
\end{equation}
where $H_u$ is a hermitian matrix, whose explicit form reads
\begin{equation}
H_u=\frac{3}{4}\left( Y_u Y_u^\dagger  - Y_d Y_d^\dagger \right) +\gamma_D \,\mathbb{I} \, ,
\end{equation}
with $\gamma_D$ involving several SM couplings. In the leading-log approximation, one has
\begin{equation}\label{eq:leading_log_approximation}
Y_u(\mu)= \left( \mathbb{I}- H_u \, \epsilon_\pi  \ln\frac{\Lambda^2}{\mu^2} \right)Y_{u}(\Lambda) \, ,
\end{equation}
which is not diagonal at the scale $ \mu \neq \Lambda $ due to the contribution of the non-diagonal down-type Yukawa matrix in $ H_u $.
The Yukawa matrix $Y_u(\mu)$, as any other matrix, can be diagonalized through a bi-unitary transformation, 
$Y_u(\mu)=U_L^u D_u (U_R^{u})^\dagger$ and thus a field redefinition, $u_{R(L)}= U_{R(L)}^{u}u_{R(L)}'$, can be used to rotate back to a basis in which $Y_u(\mu)$ is diagonal at the electroweak scale. Using unitarity, one has
\begin{equation}
U_{R}^{u\dagger} D_u^{0} \left( \mathbb{I}- H_u \, \epsilon_\pi  \ln\frac{\Lambda^2}{\mu^2} \right)^2 D_u^{0} U_{R}^u=(D_{u})^2 \, ,
\end{equation}
from which one can find~$D_{u}=D_{u}^{0}\left(\mathbb{I}+\epsilon_\pi \delta D_u+\mathcal{O}(\epsilon_\pi^2)\right)$ and $U_{R}^u=\mathbb{I}+i \epsilon_{\pi}\delta U_{R}^{u}+\mathcal{O}(\epsilon_\pi^2)$ with
\begin{equation}
i [ (D_u^{0})^2 ,\delta U_R^u ] -2 \, D_u^0 H_u D_u^0 \; \ln\frac{\Lambda^2}{\mu^2}=2 \, (D_u^{0})^2 \, \delta D_u \, .
\end{equation}
The non-diagonal elements of the equality trivially lead to (no sum over $i, j$ being considered) 
\begin{equation}\begin{aligned}
i \delta (U_{R}^{u})_{ij}&\overset{i \neq j}{=}2\,\frac{\, y_{u,i} \, H_{u,ij} \, y_{u,j}}{y_{u,i}^2-y_{u,j}^2} \ln\frac{\Lambda^2}{\mu^2} \\&= -\frac{3}{2} \sum_k \frac{ y_{u,i} \, V_{ik}y_{d,k}^2 V_{jk}^* \, y_{u,j}}{y_{u,i}^2-y_{u,j}^2} \ln\frac{\Lambda^2}{\mu^2}\, , \end{aligned}
\end{equation}
where $y_{u,i}\equiv D_{u,ii}$ and $y_{d,i}\equiv D_{d,ii}$. Thus, starting from an up-type Yukawa matrix diagonal at $\Lambda_{\mathrm{UV}}$, the back rotation needed to diagonalize back $Y_u(\Lambda_{\mathrm{EW}})$ implies
\begin{equation}
t_R  \to  -\frac{3}{2}\frac{\epsilon_\pi\,y_t \, y_u}{y_t^2-y_u^2}\ln\frac{\Lambda_\text{UV}^2}{\Lambda_\text{EW}^2} \sum_{k=d,s,b}V_{tk} \, y_k^2 \, V_{uk}^*  \,  u'_R \, , 
\end{equation}
which is Eq.~(\ref{eq:backrotation}) of the main text.
Similarly, the analogous expression for the rotation of left-handed degrees of freedom is
\begin{equation}\begin{aligned}
i \delta (U_{L}^{u})_{ij} &\overset{i \neq j}{=} H_{u,ij} \frac{y^2_{u,i} + y^2_{u,j}}{y_{u,i}^2 - y_{u,j}^2} \ln\frac{\Lambda^2}{\mu^2} \\&= - \frac{3}{4} \sum_k V_{ik} y_{d,k}^2 V_{jk}^* \frac{y^2_{u,i} + y^2_{u,j}}{y_{u,i}^2 - y_{u,j}^2} \ln\frac{\Lambda^2}{\mu^2} \, .\end{aligned}
\end{equation}

\section{Details on the derivation of bounds on top BNV}

In this appendix, we give some technical details aimed to facilitate the reproducibility of our main results. Our starting point is the branching ratios in the LEFT as obtained in Ref.~\cite{Nath:2006ut}. The translation to the LEFT basis of Ref.~\cite{Jenkins:2017dyc} is as follows 
\begin{align} \nonumber
C_{LR}^{e_\ell} &= [L_{duu}^{S,LR}]_{111\ell} \, , &
C_{RL}^{e_\ell} &= [L_{duu}^{S,RL}]_{111\ell}\, , \, \nonumber\\
C_{RR}^{e_{\ell}} &= [L_{duu}^{S,RR}]_{111\ell} \, , &
C_{LL}^{e_\ell} &= [L_{duu}^{S,LL}]_{111\ell} \, , \, \nonumber\\
C_{RL}^{\nu_\ell} &= [L_{dud}^{S,RL}]_{111\ell} \, , &
C_{LL}^{\nu_{\ell}} &= -[L_{udd}^{S,LL}]_{111\ell} \, , \, \nonumber\\
\tilde{C}_{LR}^{e_\ell} &= [L_{duu}^{S,LR}]_{211\ell} \, , &
\tilde{C}_{RL}^{e_\ell} &= [L_{duu}^{S,RL}]_{211\ell} \, , \, \nonumber\\
\tilde{C}_{RR}^{e_{\ell}} &= [L_{duu}^{S,RR}]_{211\ell} \, , &
\tilde{C}_{LL}^{e_\ell} &= [L_{duu}^{S,LL}]_{211\ell} \, , \, \nonumber\\
\tilde{C}_{RL1}^{\nu_\ell} &= [L_{dud}^{S,RL}]_{211\ell} \, , &
\tilde{C}_{RL2}^{\nu_\ell} &= [L_{dud}^{S,RL}]_{112\ell} \, , \, \nonumber\\
\tilde{C}_{LL1}^{\nu_{\ell}} &= -[L_{udd}^{S,LL}]_{121\ell} \, , &
\tilde{C}_{LL2}^{\nu_\ell} &= -[L_{udd}^{S,LL}]_{112\ell} \, .
\end{align}
Note that $[\mathcal{O}_{ddu}^{SLR}]_{1211}=-[\mathcal{O}_{ddu}^{SLR}]_{2111}$ is not included in the basis of Ref.~\cite{Nath:2006ut} and it can mediate some of the studied transitions. Nevertheless, it does not match to the SMEFT at the studied order, so we do not need to take it into account. We then perform the dimension-6 LEFT evolution to the electroweak scale
using the \texttt{Mathematica} package \texttt{DsixTools}~\cite{Fuentes-Martin:2020zaz}, which essentially implements the same $\beta$ functions of Eqs. (C.88)-(C.96) of Ref.~\cite{Jenkins:2017dyc}. This evolution plays a minor role in our analysis. The dimension-6 SMEFT-LEFT matching is done using the results of Ref.~\cite{Dekens:2019ept}. Concretely we use the analytic expressions presented in the supplementary material of that work. We use the same default inputs as the ones in \texttt{DsixTools}~\cite{Fuentes-Martin:2020zaz}, translating them to the up basis. We thus obtain the branching ratios as a function of the SMEFT Wilson coefficients at the electroweak scale in the (redundant) SMEFT basis. The redundancies are~\cite{Abbott:1980zj,Fuentes-Martin:2020zaz} 
\begin{align}\nonumber
\mathcal{Q}^{qque}_{ijkl}&=\mathcal{Q}^{qque}_{jikl}, \qquad i>j \, , \, \nonumber\\\mathcal{Q}^{qqq\ell}_{211l}&=\mathcal{Q}^{qqq\ell}_{112l} \; , \quad \mathcal{Q}^{qqq\ell}_{221l}=\mathcal{Q}^{qqq\ell}_{122l} \, , \, \nonumber\\ \mathcal{Q}^{qqq\ell}_{311l}&=\mathcal{Q}^{qqq\ell}_{113l} \; , \quad \mathcal{Q}^{qqq\ell}_{322l}=\mathcal{Q}^{qqq\ell}_{223l} \; , \; \\
\mathcal{Q}^{qqq\ell}_{33il}&=\mathcal{Q}^{qqq\ell}_{i33l} \; ,\qquad i=1,2 \, , \, \nonumber\\
\mathcal{Q}^{qqql}_{312l}&=-\mathcal{Q}^{qqql}_{132l}+\mathcal{Q}^{qqql}_{213l}+\mathcal{Q}^{qqql}_{231l} \, , \, \nonumber\\
\mathcal{Q}^{qqql}_{321l}&=-\mathcal{Q}^{qqql}_{231l}+\mathcal{Q}^{qqql}_{123l}+\mathcal{Q}^{qqql}_{132l} \, . \label{eq:equivoperators}
\end{align}
Whenever there is a relation $\mathcal{Q}_1=\sum_{i=2}^n \mathcal{Q}_{i}$, any observable calculated in that redundant basis $\Gamma=\Gamma(\sum_{i=1}^n a_{i}^{\Gamma} C_i+\cdots)$ (where the ellipsis denotes the contributions from yet other operators) is such that $a_{1}^{\Gamma}=\sum_{i=2}^n a_{i}^{\Gamma}$, since the $a_{i}^{\Gamma}$ factors come from matrix elements of the corresponding operators. Our choice to remove the redundancy, natural within the \texttt{DsixTools} interface, is taking $C_{1}=\sum_{i=2}^{n}C_i$,
with the ordering provided in Eq.~(\ref{eq:equivoperators}), i.e., these relations are used to replace the operators on the left-hand side by those on the right-hand side.
With this identification one finds the correct result for $\Gamma$
for the full basis within the chosen convention, $C_{1}=\sum_{i=2}^{n}C_i$. Thus, in a one-parameter-at-a time fit to $C_{123l}^{qqql}$, one obtains the experimental bound on the Wilson coefficient $C_{123l}^{qqql}$ assuming by convention that the Lagrangian set-up is $\mathcal{L}=C_{123l}^{qqql}\, \mathcal{O}_{123l}^{qqql}+C_{123l}^{qqql}\, \mathcal{O}_{321l}^{qqql}$.
This is because the operator $ \mathcal{Q}^{qqql}_{123l} $ appears in the right-hand side of the substitution shown in Eq.~\eqref{eq:equivoperators} for the operator $\mathcal{Q}^{qqql}_{321l}$.
The identifications of the Wilson coefficients above must be consistently implemented in the $\beta$ functions of the SMEFT running.

Taking this into account, we numerically solve the RGEs for the dimension-4 SM couplings using the \texttt{DsixTools} default SM inputs in the flavor basis in which the up Yukawa is diagonal at the electroweak scale. Using the obtained dimension-4 running couplings, we then numerically solve the RGEs corresponding to the dimension-6 BNV Wilson coefficients. Special care needs to be taken not to miss the numerically small mixings involving Yukawas, relevant for our analysis. This gives us the branching ratios as a function of the Wilson coefficients at the TeV scale, which we use for the fits. The results are displayed in Fig.~\ref{fig:topBNV_bounds}.

\bibliographystyle{apsrev4-1}
\bibliography{bibliography}

\begin{thebibliography}{67}
\bibitem [{foo()}]{footnote}%
  \BibitemOpen
  \href@noop {} {\bibinfo  {journal} {Direct bounds on BNV in bottom-meson decays could be used to constrain operators carrying third generation quark indices, which however are not competitive with respect to the indirect constraints discussed in the main text}\ }\BibitemShut {NoStop}%
\end{thebibliography}

\end{document}